\documentclass[fleqn,usenatbib]{mnras}

\usepackage{newtxtext,newtxmath}

\usepackage[T1]{fontenc}
\usepackage{comment}

\usepackage{subcaption}

\DeclareRobustCommand{\VAN}[3]{#2}
\let\VANthebibliography\thebibliography
\def\thebibliography{\DeclareRobustCommand{\VAN}[3]{##3}\VANthebibliography}

\usepackage{graphicx}	
\usepackage{amsmath}	

\newcommand{\Duo}{{\sc Duo}}

\newcommand{\cm}{cm$^{-1}$}

\newcommand{\ai}{\textit{ab initio}}

\newcommand{\X}{X~$^2\Pi$} 
\newcommand{\A}{A~$^2\Sigma^+$}

\title[Exomol line list - NaO]{Exomol line list - NaO}
\title[ExoMol XLIII: NaO  line list]{ExoMol molecular line lists -- XLIII: Rovibronic transitions corresponding to the close-lying X~$^2\Pi$ and A~$^2\Sigma^+$ states of NaO}

\author[G. B. Mitev et al.]{
G. B. Mitev,$^{1}$
S. Taylor,$^{1}$
Jonathan Tennyson,$^{1}$
S. N. Yurchenko,$^{1}$
A. A. Buchachenko,$^{2}$
A. V. Stolyarov$^{3}$
\\
$^{1}$ Department of Physics and Astronomy, University College London, Gower St,  London WC1E 6BT, UK\\
$^{2}$ Skolkovo Institute of Science and Technology, Skolkovo Innovation Center, Moscow  121205,  Russia\\
$^{3}$ Department of Chemistry, Lomonosov Moscow State University, Moscow, 119991, Russia}

\date{Accepted XXX. Received YYY; in original form ZZZ}

\pubyear{YYYY}

\begin{document}
\label{firstpage}
\pagerange{\pageref{firstpage}--\pageref{lastpage}}

\maketitle

\begin{abstract}
The sodium monoxide radical (NaO) is observed in night glow  in the Earth's mesosphere and likely has astronomical importance. This study concerns the optical transitions within the ground X~$^2\Pi$ state and to the very low-lying ($T_{\rm e}\approx 2000$ cm$^{-1}$)  excited  A~$^2\Sigma^+$ state. A line list consisting of rovibronic term values, allowed electric dipole transitions, Einstein coefficients, and partition functions for varying temperature are produced due to a variational solution of the coupled-channel Schr\"{o}dinger equations using the program \Duo. MRCI \textit{ab initio} calculations characterising the potential energy curves of the both states, spin-orbit  and $L$-uncoupling non-adiabatic matrix elements, as well as permanent and transition dipole moments were integral in formation of the final deperturbation model. \textit{Ab initio} potential energy curves are represented in the analytical Extended-Morse-Oscillator form and refined, along with the spin-orbit and $L$-uncoupling functions, by least-squares fitting to the available spectroscopic data. The input experimental data consisted of pure rotational transitions within the fine structure components of the X~$^2\Pi$ state for $v''\in [0,3]$ vibrational levels as well as the rovibronic A~$^2\Sigma^+(v'=0) \leftarrow$ X~$^2\Pi(v''=0)$ transitions, both with limited coverage over rotational excitation. The lack of data detailing the vibrational structure of the X and A states prompts a request for further experimental study of higher excited levels which would provide a robust spectroscopic model. The NaO NaOUCMe line list is available via www.exomol.com  and the CDS database.
\end{abstract}




\section{Introduction}

Air-glow, sometimes referred to as night-glow, was first reported by \citet{29Slipher} and has been identified as the emission of Na D-lines. It is proposed that these lines
arise from  the Chapman mechanism 
\citep{39Chapman,10SaSeChSr,12PlOeMiSa} with some modifications suggested  in light of modern measurements \citep{05SlCoHuSa,10SaSeChSr}, as given by 
\begin{equation}\label{eq:Chap1}
    \textrm{Na} + \textrm{O}_3 \xrightarrow{k_1} \textrm{NaO} + \textrm{O}_2
\end{equation}
\begin{align}\label{eq:Chap2}
    \textrm{NaO} + \textrm{O} & \xrightarrow{\alpha k_2} \textrm{Na}(^2P) + \textrm{O}_2\\
                              & \xrightarrow{(1-\alpha)k_2} \textrm{Na}(^2S)+\textrm{O}_2
\end{align}
\begin{equation}\label{eq:Chap3}
    \textrm{Na} (^2P) \rightarrow \textrm{Na} (^2S) + h\nu (\textrm{D}_1,\textrm{D}_2),
\end{equation}
where $k_{1,2}$ are the temperature-dependent rate constants, $\alpha$ is the branching ratio, and $\textrm{D}_{1,2}$ are the D-line transition energies.

This mechanism suggests that Na atoms, which are assumed to be ablated from meteors, react with O$_3$ in the atmosphere to form NaO, re-reacting with O atom to reform Na. Long lasting visible meteor trails have been attributed to the production of electronically excited Na by in the said reaction chain \citep{76KoEl,99JoWoKo.NaO}. In the Chapman mechanism, the rate limiting reaction is the formation of NaO, see Eq.~\eqref{eq:Chap1}. As this reaction takes place in the mesosphere, the variation of mesospheric ozone concentration affects Na air-glow intensity. As altitude profiles of ozone are available, ozone concentration variation can be used for deeper study into Sodium air-glow and subsequently NaO. \citep{03Plane,10SaSeChSr}. 

The experimental spectroscopic data related to optical transitions of the gas-phase NaO molecules is very sparse. \citet{89YaFuHi.NaO} provide comprehensive microwave data on the pure rotational transitions of the ground X$^2\Pi$ electronic state for the lowest vibrational $v''=0,1,2,3$ levels. However, no experimental data on ro-vibrational transitions are available, and hence, the separations between vibrational levels are not yet experimentally determined. Some information is available on the B~$^2\Pi\rightarrow$~\X\ emission spectrum, however this is minimal, unassigned with no characterisation of the B state \citep{84PfGoxx.NaO, 96PuShWi.NaO}, which sometimes is assigned as C~$^2\Pi$ state \citep{91LaPaBa.NaO}. The only experimental observation of the lowest excited A~$^2\Sigma^+$ state is due to \citet{99JoWoKo.NaO} who could assign a handful of the low rotational levels $J = [4.5,19.5]$ belonging to the rovibronic A~$^2\Sigma^+(v'=0) -$ X~$^2\Pi(v''=0)$ transitions.

The electronic structure of NaO demonstrates  pronounced ionic inter-atomic bond leading to large permanent dipoles as well as the close-lying ground X~$^2\Pi$ and first excited A~$^2\Sigma^+$ state converging to different dissociation thresholds. The low excitation energy of the A~$^2\Sigma^+$ state means that its spin-allowed A--X electronic transition lies in the infrared, largely in the 3 to 5 $\mu$m region. From the theoretical viewpoint, the A and X states are expected to be strongly coupled to each other due to spin-orbit and electronic-rotational intramolecular perturbations. \textit{Ab initio} potential energy curves (PECs) and permanent dipole moments are available for both lowest electronic states \citep{09BuLi,91LaPaBa.NaO,99SoLeGaWr.NaO}. However, the corresponding non-adiabatic coupling matrix elements are not known so far. 

In this paper we present new \ai\ PECs for both X~$^2\Pi$ and A~$^2\Sigma^+$ states of NaO accompanied by spin-orbit and $L$-uncoupling non-adiabatic matrix elements which we tune to the limited available spectroscopic data. Finally,  a rovibronic line list for NaO is simulated to be suitable for modeling spectra up to temperature 2500~K. This line list is constructed as part of the ExoMol project \citep{jt528}. We note that a previous ExoMol line list for CaO \citep{jt618}, constructed using similar methodology to that employed here, was used by \citet{18BeBoSa.CaO} to assign a spectrum of CaO in the wake of the Beneshov bolide at the height of 29 km above the Earth's surface.

\section{Method}

\subsection{\textit{Ab initio} electronic structure calculation} \label{sec:abinitio}

The initial set of required potential energy curves (PECs), permanent and transition dipole moments for the ground X$^2\Pi$ and near-lying excited A$^2\Sigma^+$ states of NaO were obtained in the framework of \ai\ electronic structure calculations which were accompanied by the evaluation of the relevant non-adiabatic matrix elements for the spin-orbit and Coriolis (electron-rotational) coupling.  Scalar relativistic effects are introduced by means of the second-order Douglas-Kroll Hamiltonian \citep{Reiher04}. All calculations were performed in the range of internuclear distance $R\in [1.0,5.0]$~\AA~using the MOLPRO program suite~\citep{MOLPRO2010}.

For both Na and O atoms, aug-cc-pVQZ all electrons basis sets was used. The optimized molecular orbitals (MO’s) were generated using the state-averaged self-consistent field (SA-CASSCF) method ~\citep{Werner85}, taking the lowest doublet (1-3)$\Pi$, (1)$\Sigma^+$, (1)$\Sigma^-$, and (1)$\Delta$ states with equal weights. Internally-contracted multi-reference configuration interaction (MR-CISD) calculations~\citep{Knowles92} followed, in which impact of higher excitation on the correlation energy was taken into account implicitly using the Davidson correction \citep{Langhoff74}. For both SA-CASSCF and MR-CISD steps the active space consisted of $7a_1$, $3b_1$ and $3b_2$ orbitals corresponding to the point group symmetry $C_{2v}$, while the lowest two $a_1$ orbitals were kept to be doubly occupied. 
The full counterpoise correction \citep{Boys70} was accounted for each state individually with the residual size-consistency error eliminated at $R = 20$ {\AA}. 

Both permanent and transition dipole moments as well as non-adiabatic $L$-uncoupling matrix element were computed using the MR-CISD wave functions. The same wave functions and the full Breit-Pauli operator were used to compute the SOCs ~\citep{Berning2000}.

\subsection{Fitting procedure} \label{sec:Fitting}

The adiabatic PECs \textit{ab initio} calculated above in Sec.~\ref{sec:abinitio} were combined with the relevant spin-orbit coupling (SOC) and $L$-uncoupling  functions in order to evaluate non-adiabatic rovibronic term values and corresponding multi-channel wavefunctions for the mutually perturbed X~$^2\Pi$ and A~$^2\Sigma^+$ states of NaO. However, the pure \textit{ab initio} estimates are \textit{a priori} and are not accurate enough to predict the rovibronic energy with the spectroscopic accuracy required. For better agreement between predicted and observed transitions, PECs and coupling curves are refined by making constrained adjustments to the parameters which describe them (see below). This is done by a least-squares fitting to experimental data and is performed in Duo, an open-source Fortran 2009 program which provides variational solutions to the coupled rovibronic Schrodinger equations for a general open-shell diatomic molecule \citep{jt609}.

The input set of the experimental data taken from  \citep{89YaFuHi.NaO} consisted of pure rotational transition frequencies (with $\Delta v_X = 0$ and $v_X = 0,1,2,3$) within  up to $J'' = 16.5$. \citet{99JoWoKo.NaO} provided rovibronic frequencies for the A -- X $(0,0)$ band with up to $J'' = 15.5$ together with relative absorption intensities. Joo {\it et al.} also observed the $v^{\prime} - v^{\prime\prime} = 1$ band suggesting an A~$^2\Sigma^+$ state vibrational spacing of 498.9~cm$^{-1}$. This spacing was included in the present fit using an artificial $A(v' = 1) - A(v'' = 0)$ line for $J' = J'' = 0.5$.

The \textit{ab initio} point-wise PECs of both states treated were approximated using the fully analytical Extended Morse Oscillator (EMO) form~\citep{EMO,jt609,dPotFit}: 
\begin{eqnarray}\label{eq:EMO}
    V_{\textrm{EMO}}(r) = T_{\rm e} + D_{\rm e}\left[1 - e^{-\beta(r)(r-r_{\rm e})}\right ]^2
\end{eqnarray}    
where $T_{\rm e}$ is the electronic term, $D_{\rm e}$ is the dissociation energy and $r_{\rm e}$ is the equilibrium distance. In contrast to the conventional Morse potential, the $r$-dependent exponent coefficient $\beta(r)$ in Eq.~\eqref{eq:EMO} is defined as the polynomial series
\begin{eqnarray}\label{eq:beta}
     \beta(r) = \sum^{N}_{i=0}\beta_i [y_p(r)]^i;\quad
         N =
    \begin{cases}
    N_+ & \textrm{for} ~~ r > r_{\rm e} \\
    N_- & \textrm{for} ~~ r \leq r_{\rm e} 
    \end{cases}
\end{eqnarray}   
with respect to the reduced coordinate
\begin{eqnarray}
\label{eq:surkus}
    y_p(r) = \frac{r^{p}-r_{\rm e}^{p}}{r^{p}+r_{\rm e}^{p}},
\end{eqnarray}
which was first introduced by \citet{84SuRaBo.method}. 

At the first step, the EMO potentials \eqref{eq:EMO} were  least-squares fit to the original \textit{ab initio} PECs in Python.  The electronic energy of the ground state was fixed to zero, and both PECs were fit to fourth order in the $\beta(r)$ polynomial expansion \eqref{eq:beta}. The parameter $p$ in Eq.\eqref{eq:surkus} was fixed to 6 and 4 for the ground and excited state, respectively.

At the second step,  the resulting \ai\ EMO PECs were refined by adjusting to the experimental transition frequencies \citep{99JoWoKo.NaO, 89YaFuHi.NaO} where only $\beta_0$(X, A), $T_{\rm e}$(A), $r_{\rm e}$(X) and $r_{\rm e}$(A) fitting parameters were varied. Furthermore, both diagonal, $A_{\rm SO}(r)$, and off-diagonal, $\xi_{\textrm{\rm AX}}(r)$, spin-orbit coupling  (SOCs) functions as well as $L$-uncoupling  function, $L_{\textrm{\rm AX}}(r)$, between \A\ and \X\ states were also fitted simultaneously with the trial EMO PECs above. This was particularly done using the  facility in \Duo\ to adjust curves using a \textit{morphing} function, $f_{\rm{C}}$:
\begin{eqnarray}\label{DUOmorph}
    f_{\textrm{C}}^{\textrm{morphed}}(r) = f_{\rm C}(r)f^{ab}_{\textrm{c}}(r),
\end{eqnarray}
where {\rm C} is the coupling curve in question.  $f^{ab}_{\textrm{c}}(r)$ are the original \textit{ab initio} point-wise functions  interpolated by ordinary cubic splines with so-called natural boundary conditions. The morphing function, $f_{\rm C}(r)$, in Eq.(\ref{DUOmorph}), is tacitly assumed to be an analytical function of the internuclear distance $r$, and it was modelled by the following linear function
\begin{eqnarray}\label{eq:PD24}
    f_{\rm C}(r) = B_0(1-y_p(r))+y_p(r) B_{\infty}
\end{eqnarray}
of the \v{S}urkus-like variable, $y_p(r)$, (see Eq.~\eqref{eq:surkus}), where the equilibrium distance $r_{\rm e}$ is substituted for the fixed parameter $r_0=2.053$~\AA~. The parameter $p=6$ and $B_{\infty}=1$ are also fixed for all morphing functions while only $B_0$ parameter is kept to be variable.

Morphing (Eq: \ref{DUOmorph}) was used  as it avoids the need for an analytical representation  of the \textit{ab initio} point-wise coupling curves by changing their original shapes directly. This approach is, however, not normally applied for PECs for the reason of difficulty to control the equilibrium parameter $r_{\rm e}$, which is a value of interest not explicitly derivable from this morphing form.

\section{Results and Discussion}\label{sec:Results}
\subsection{Potential energy curves, spin-orbit and $L$-uncoupling  matrix elements}

\begin{figure}
     \centering
     \begin{subfigure}[b]{0.4\textwidth}
         \centering
         \includegraphics[width=\textwidth]{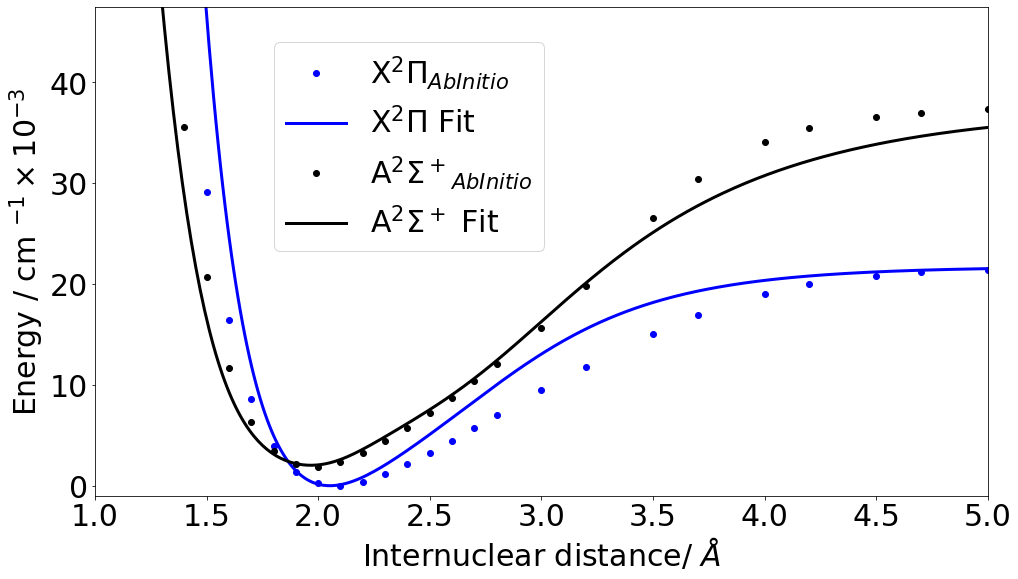}
     \end{subfigure}
     \hfill
     \begin{subfigure}[b]{0.4\textwidth}
         \centering
         \includegraphics[width=\textwidth]{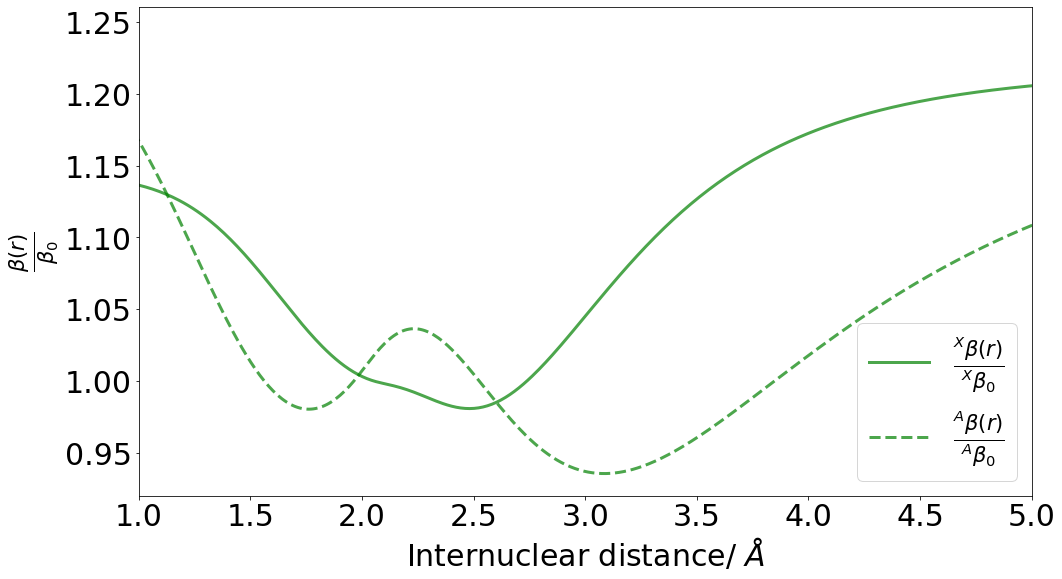}
     \end{subfigure}
        \caption{Empirical EMO (fitted) and original {\it ab initio} point-wise potential energy curves obtained for the X~$^2\Pi$ and A~$^2\Sigma^+$ states of NaO (upper panel);
        Distance dependent exponent coefficient for X~$^2\Pi$ and A~$^2\Sigma^+$ state EMO fits as normalised by $\beta_0=A_0$ (lower panel).}
        \label{fig:FinalPECs}
\end{figure}

\begin{figure}
     \centering
     \begin{subfigure}[b]{0.4\textwidth}
         \centering
         \includegraphics[width=\textwidth]{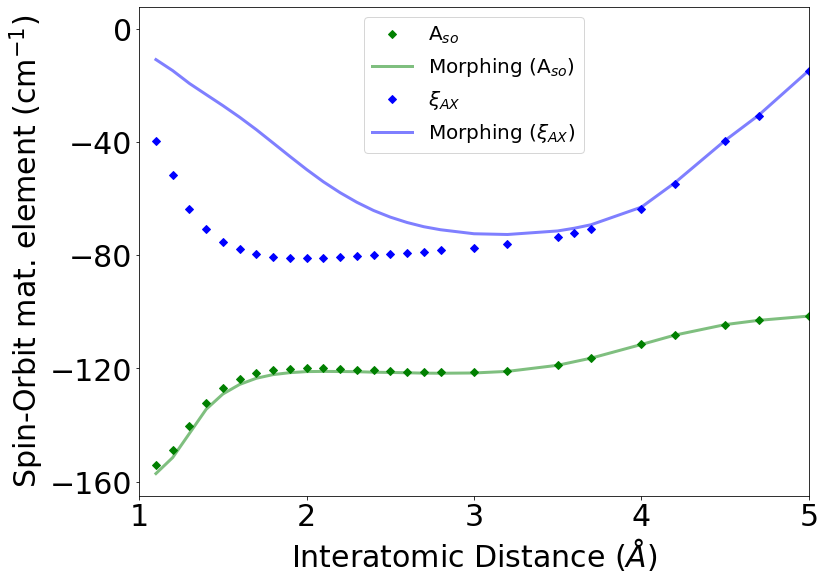}
     \end{subfigure}
     \hfill
     \begin{subfigure}[b]{0.4\textwidth}
         \centering
         \includegraphics[width=\textwidth]{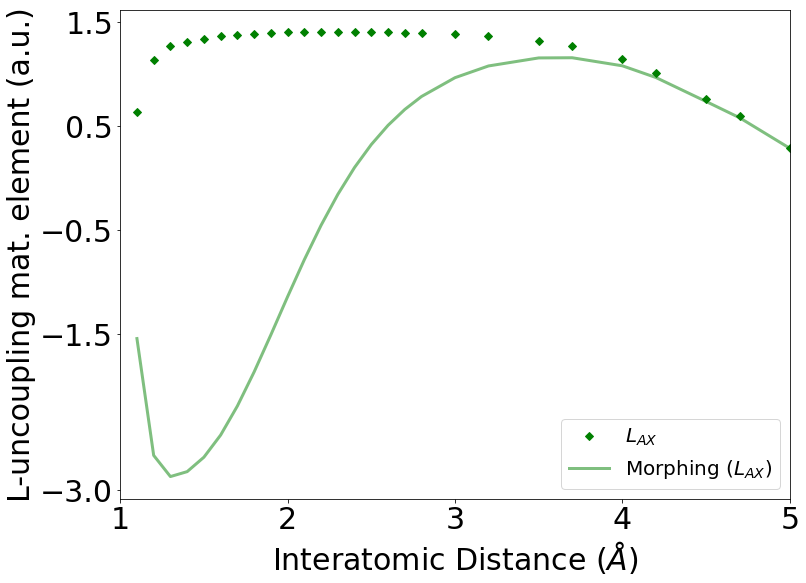}
     \end{subfigure}
        \caption{Upper: Diagonal, $A_{\rm SO}(r)$, and off-diagonal, $\xi_{\rm AX}(r)$, spin-orbit coupling curves \textit{ab initio} calculated for the X-state and A-states. Lower: $L_{\rm AX}(r)$ is the corresponding non-adiabatic matrix element of electron-rotational interaction between the X-state and A-state.}
    \label{fig:SO_Lc_NaO}
\end{figure}

The resulting \textit{ab initio} adiabatic PECs obtained for the X- and A-states of NaO are depicted on Fig.~\ref{fig:FinalPECs} together with their empirical (adjusted) counterparts. The fitted parameters of the EMO potentials (Eq.~\eqref{eq:EMO}) for the both states treated are collected in Table~\ref{Table:AXParam}. The corresponding \textit{ab initio} diagonal spin-orbit  function, $A_{\rm SO}(r)$, of the ground X-state as well as the off-diagonal SOC, $\xi_{\rm AX}(r)$, and the $L$-uncoupling, $L_{\rm AX}(r)$, functions are shown in Fig.~ \ref{fig:SO_Lc_NaO}. The relevant morphing parameters for these functions are given in Table~\ref{Table:AXParam}.

The fitted EMO PECs and 
coupling functions reproduce the input experimental data with an overall root-mean-squares (RMS) error of  $\approx 0.086~\textrm{cm}^{-1}$. This RMS is deceptively large due to the pronounced discrepancy of the A~$^2\Sigma^+$ artificial line, skewing the mean RMS, see Table \ref{Table:RMS}. Indeed, for the ground X~$^2\Pi$ state, there is agreement with the experiment on order of $\sim 0.006$ \cm and for the $\rm{A}(v' =0) \leftarrow \rm{X}(v'' = 0)$ case there is agreement on order $\sim 0.02$ \cm.  

The \textit{ab initio} $L_{\rm AX}(r)$ function obeys Van Vleck's hypothesis~\citep{Vleck} of \emph{pure precession} in the interval $r\in [1.3,3.7]$~\AA~:
\begin{eqnarray}\label{VanVleck}
L_{\rm AX}(r)\approx\sqrt{L(L+1)- |\Lambda|(|\Lambda|\pm 1)}=\sqrt{2},
\end{eqnarray}
where the total electronic angular momentum of molecule $L$ is equal to 1 while its projection on the internuclear axis is zero.   
Moreover, the relationship
\begin{eqnarray}\label{VanVleck1}
\xi_{\rm AX}(r)\approx A_{\rm SO}(r)L_{ AX}(r)/2
\approx A_{\rm SO}(r)/\sqrt{2}
\end{eqnarray}
between diagonal and off-diagonal spin-orbit functions is valid as well. 

The \textit{ab initio} SO splitting of the ground X-state is found to be close to the empirical $A_{\rm SO}$ value determined by \citet{89YaFuHi.NaO} (see Table~\ref{Table:LambdaParam}). The reliability of the \textit{ab initio} off-diagonal spin-orbit $L$-uncoupling functions is indirectly confirmed by a good agreement of the $\Lambda$-doubling constants ($p$, $q$) empirically obtained for the regular perturbed levels of the ground state by \citet{89YaFuHi.NaO} with their theoretical counterparts roughly estimated for the lowest vibrational levels, $v_X$, according to the relations ~\citep{field, PupyshevCPL94}:
\begin{eqnarray}\label{p}
p = 2\langle v_X|B\xi_{\rm AX}L_{\rm AX}/\Delta U_{XA}|v_X\rangle
\end{eqnarray}
\begin{eqnarray}\label{q}
q = 2\langle v_X|[BL_{\rm AX}]^2/\Delta U_{XA}|v_X\rangle
\end{eqnarray}
where $B=1/2\mu R^2$, $\mu$ is the reduced molecular mass, $\Delta U_{XA}=U_X-U_A$ is the difference of PECs, while $|v_X\rangle_R$ are the vibrational wavefunctions of the ground state.

\begin{table}
\centering
\caption{Root-mean squares errors $\Delta$(RMS) in \cm\ comparing fitted model to a: \citet{89YaFuHi.NaO} and b: \citet{99JoWoKo.NaO} rotational transitions.}
\label{Table:RMS}
\begin{tabular}{lll}
\hline\hline
$v$ & $\Delta_{\textrm{a}}$ & $\Delta_{\textrm{b}}$         \\
\hline
0   & 0.002664              & 0.022501                      \\
1   & 0.002846              & 0.679901$^a$            \\
2   & 0.002758              &                               \\
3   & 0.017115              &                               \\
Ave.& 0.006345              & 0.351201                      \\
\hline
\end{tabular}
\flushleft{
$^a$: artificial line, see Sec. \ref{sec:Fitting}.}
\end{table}

\begin{table}
\centering
\caption{PEC and morphing parameters obtained for the \X\  and \A\ states of NaO. For the definition of the EMO parameters see Eq.~(\ref{eq:EMO}), and for morphing  parameters see Eq.~(\ref{eq:PD24}).\\}
\label{Table:AXParam}
\begin{tabular}{lll}
\hline\hline
\multicolumn{3}{c}{PEC Parameterisation}                        \\
\hline
EMO Param.   & X $^2\Pi$             & A $^2\Sigma^+$        \\
\hline
$T_{\rm e}$     & 0                     & 2029.147$^a$  \\
$r_{\rm e}$     & 2.05189$^a$     & 1.96723$^a$  \\
$D_{\rm e}$     & 21735.7               & 35496.4               \\
${p}$           & 6                     & 4                     \\
${N}_-^{b}$         & 2                     & 2                     \\
${N}_+^{b}$         & 4                     & 4                     \\
${\beta}_0$     &  1.50708327083484$^a$  &  0.989061691136872$^a$  \\
${\beta}_1$     & -0.04809897091909            &  0.181504940347642  \\
${\beta}_2$     &  0.16743487156786            &  0.424100841906937  \\
${\beta}_3$     & -0.95627676523677            & -2.926458322108940 \\
${\beta}_4$     &  1.16624139895144            &  2.522897112151440  \\
\hline
\multicolumn{3}{c}{$B_0^a$ parameters used for morphed coupling functions$^c$}                        \\
\hline
Function                 & $B_0$                   \\
\hline
$A_{\textrm{so}}$        & 1.0098\\
$\xi_{\textrm{AX}}$      & 0.6277\\
$L_{\textrm{AX}}$        & -0.7455\\
\hline
\end{tabular}
\\
\flushleft{
$^a$: value acheived via varying in Duo. All others were fixed.\\
$^b$: As in Eq. \eqref{eq:EMO}, the $N_{-}$ and $N_{+}$ values indicate the limits to which the expansion is summed depending on whether $r < r_{\rm e}$ or $r > r_{\rm e}$ respectively.\\
$^c$:For Morphing parameters: As per Sec \ref{sec:Fitting}, $r_0 = 2.05$~\AA, $p = 6$, $B_{\infty} = 1$ (standard when using Morphing) for all three $f_C$. }

\end{table}

\begin{table}
\centering
\caption{Comparison of \textit{ab initio} calculated spin-orbit splitting ($A_{\textrm{so}}$) and $\Lambda$-doubling constants ($\textrm{p}$, $\textrm{q}$) of the ground NaO state with their empirical counterparts taken from \citet{89YaFuHi.NaO}. All values in \textrm{cm}$^{-1}$.} 
\label{Table:LambdaParam}
\begin{tabular}{ccc}
\hline\hline
Parameter & Empirical & \textit{Ab initio}\\
\hline
$A_{\textrm{so}}$ & -107 & -120\\
$\textrm{p}$ & +0.044 & +0.04\\
$\textrm{q}$ & +0.0003 & +0.0003\\
\hline
\end{tabular}
\end{table}

\subsection{Permanent and Transition Dipole Moments}

\begin{figure}
	\includegraphics[width=\columnwidth]{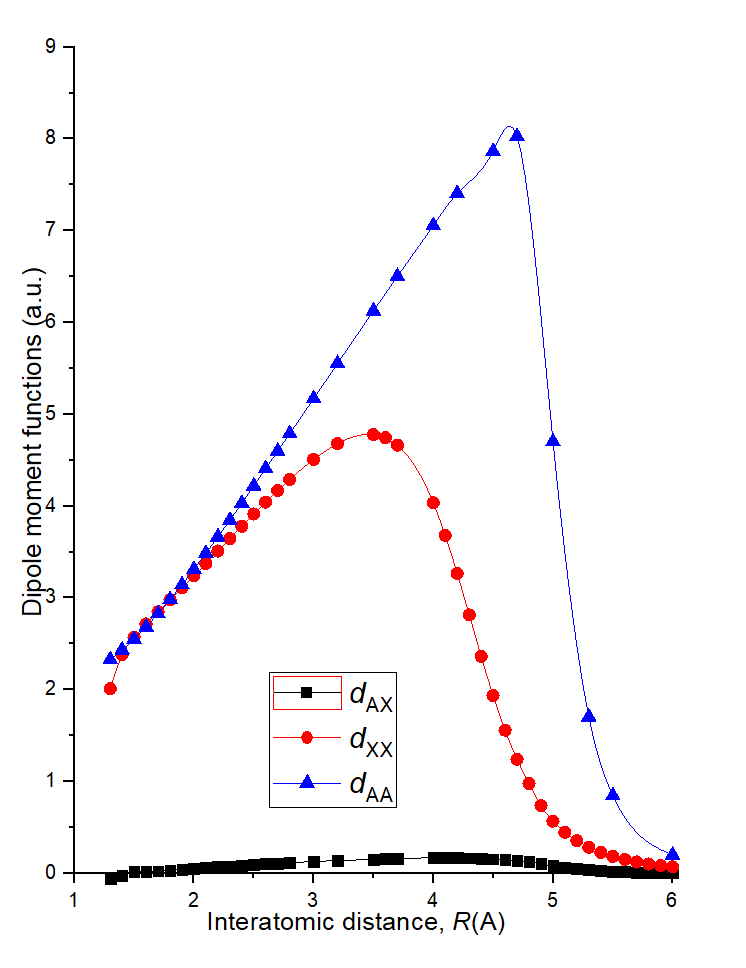}
    \caption{Permanent, $d_{\rm XX}(r)$, $d_{\rm AA}(r)$, and transition, $d_{\rm AX}(r)$, dipole moment functions \textit{ab initio} calculated for the X-state and A-states of NaO molecule.}
    \label{fig:dipole_NaO}
\end{figure}

The \textit{ab initio} permanent dipole moments $d_{\rm XX}(r)$, $d_{\rm AA}(r)$ calculated for the X$^2\Pi$ and A$^2\Sigma^+$ states of the NaO molecule are given in Fig.~\ref{fig:dipole_NaO} together with the corresponding A--X transition dipole moment $d_{\rm AX}(r)$. The present dipole moments are in a good agreement with their previous theoretical counterparts computed by \citet{91LaPaBa.NaO}. 

According to the expected ion-pair character of the interatomic bond, the permanent dipole moments of both A and X states are rapidly increasing linear functions of $r$ while the corresponding A--X electronic transition dipole moment is very small, but not negligible (see Fig.~\ref{fig:dipole_NaO}). The maximal values of the present A--X transition moment are about 0.1-0.15 a.u. which agree well with the estimate $|d_{\rm AX}|<0.2$~D given in \citet{99JoWoKo.NaO}. At $r> 3.5$~\AA, the ground state permanent dipole decreases due to the avoided crossing effect with the excited $C$~$^2\Pi$ state. An even more abrupt drop in the permanent dipole is observed for the A-state but at larger internuclear distances,
presumably also due to an avoided crossing.

\citet{72OHWaAr} give an equilibrium dipole moment for the X~$^2\Pi$ state of 8.71~D which was adopted by the JPL Molecular Spectroscopy Database \citep{JPLDatabase} to provide transition intensities. Our {\it ab initio} equilibrium value is $|d_{\rm XX}(r_{\rm e})|=8.43$~D, while the vibrationally averaged dipole moment extracted from the \Duo\ the matrix element $|\langle v_X=0|d_{\rm XX}|v_X=0\rangle|$  is  8.44~D,  which is very similar to the equilibrium dipole moment. Our results suggest that JPL predicted transitions are slightly (about 6\%) too strong.

\subsection{Partition Function}

\begin{figure}
    \includegraphics[width=\columnwidth]{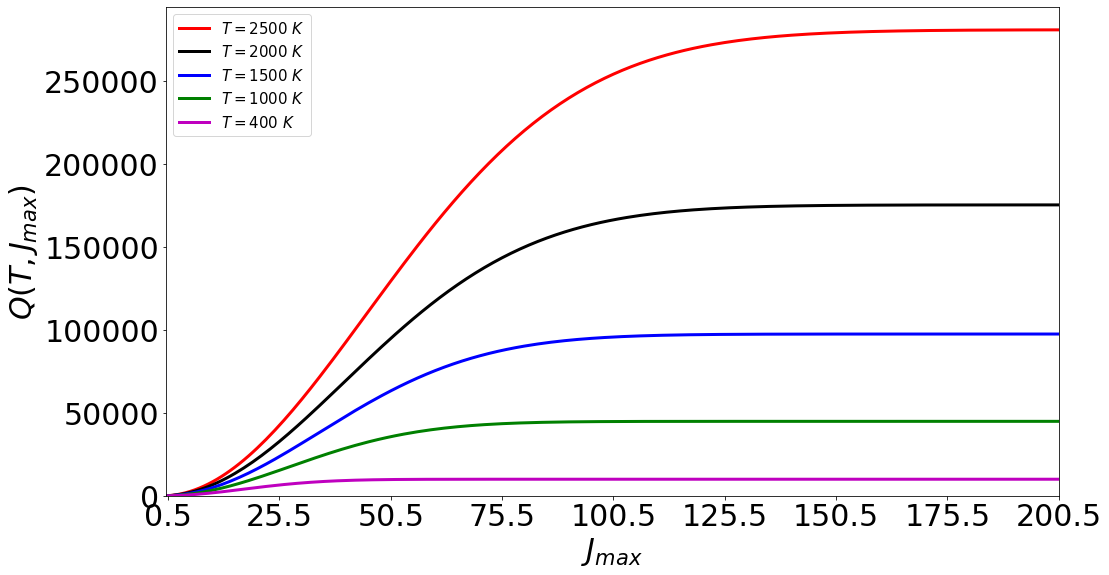}
    \caption{Partition function $Q(T)$ convergence of NaO as dependent on maximum value of rotational quantum number, $J_{\textrm{max}}$ for temperatures up to 2500~K.}
    \label{fig:PF}
\end{figure}

The temperature dependent partition function for NaO molecule was calculated as the sum:
\begin{eqnarray}\label{eq:Q}
    Q(T) = \sum_i g_i e^{\frac{-E_i}{k_BT}}, 
\end{eqnarray}
where $g_i = g_{\rm ns}(2J+1)$ is the total degeneracy of the state, $g_{\rm ns}$ is the nuclear statistical weight factor, and $k_B$ is the Boltzmann constant. 
As $Q(T)$ sums over $J$, the value of $Q(T)$ at some $T$ depends on the value of $J_{\textrm{max}}$ used in the sum. To ensure completeness of the line list, it is a requirement that, up to a given temperature, in the case of this study, 2500~K, the value of $Q(2500)$ converges with $J_{\textrm{max}}$. As shown in Fig.~\ref{fig:PF},
use $J_{\textrm{max}} = 200.5$  converges $Q$ at $=2500$~K.

\begin{figure}
     \centering
     \begin{subfigure}[b]{0.4\textwidth}
         \centering
         \includegraphics[width=\textwidth]{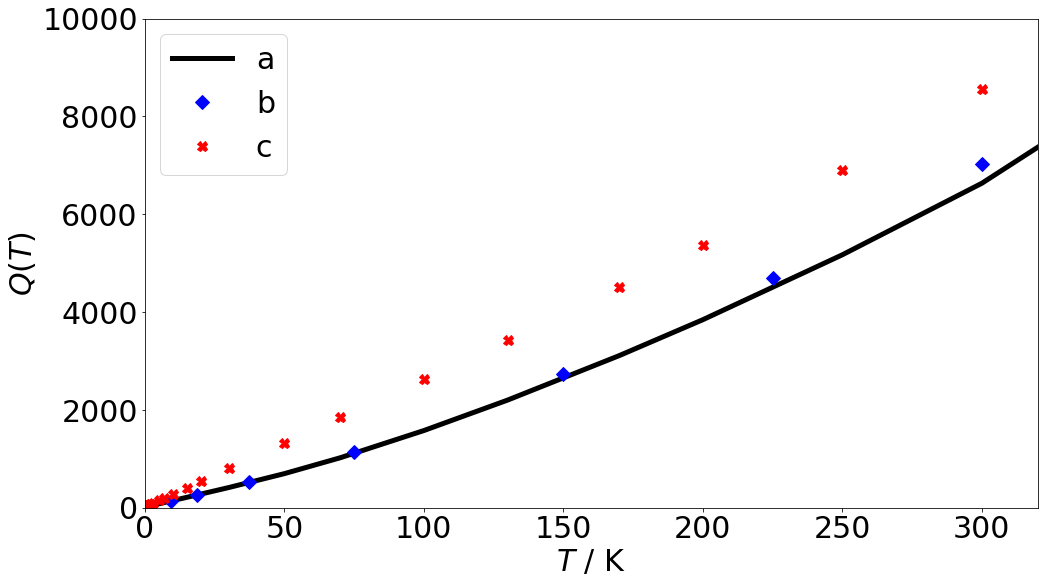}
     \end{subfigure}
     \hfill
     \begin{subfigure}[b]{0.4\textwidth}
         \centering
         \includegraphics[width=\textwidth]{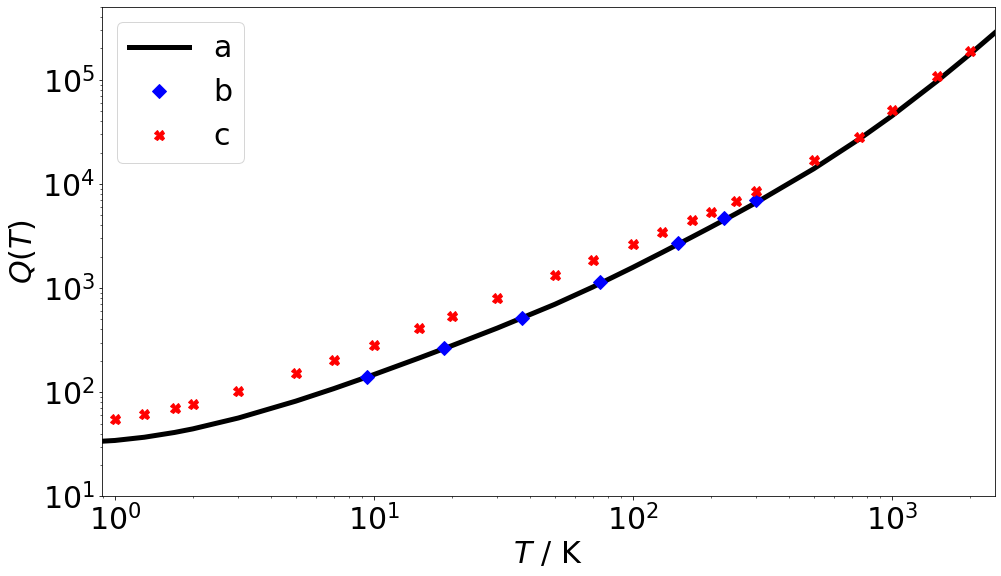}
     \end{subfigure}
        \caption{Comparison between, (a): calculated partition function based on Duo states and available data for the partition function of NaO (b): \citet{JPLDatabase}, (c): \citet{16BaCoxx.partfunc}. Note, $Q(T)$ by \citet{16BaCoxx.partfunc}  has been multiplied by four to account for the convention of including the hyperfine structure degeneration number $g_{\rm{ns}}$ in the partition function calculation, see Eq. \ref{eq:Q}. Upper: Close-up of calculated and reference partition functions for $T\in [0,300]$~K. Lower: Log scaled representation of calculated and reference partition functions for $T\in [0,2500]$~K.}
        \label{fig:PF_Total}
\end{figure}

The JPL database provides partition function values up to $T$=300~K. \citet{16BaCoxx.partfunc} provide numerical values for the partition functions of NaO up to 10~000 K which we multiply by four to account for different treatments
of the Na nuclear spin degeneracy; we (and JPL) include  $g_{\rm ns} = 2I_{\rm Na}+1=4$ (which is appropriate for $I_{\rm Na}=\frac{3}{2}$) into $Q(T)$, while \citet{16BaCoxx.partfunc} does not. A comparison between these partition functions is given in Fig.~\ref{fig:PF_Total}. The JPL values agree well with ours (see Table \ref{Table:PF}), with level of agreement decreasing as temperature goes up which likely due to a lack of higher vibrational levels contribution in JPL partition sum. However, there is  poor coincidence with \citet{16BaCoxx.partfunc} at low temperature with agreement improving for higher temperature. The reason for the low temperature disagreement between \citet{16BaCoxx.partfunc} and the results of JPL and our own is unclear but we believe our values are correct in this region. It should be noted that the data from \citet{16BaCoxx.partfunc} is based on calculated rather than experimental molecular constants taken from \citet{79HeHuxx.book}.

\begin{table}
\centering
\caption{Partition Function, $Q(T)$ as calculated using the Duo computed energies compared to partition functions from JPL database.}
\label{Table:PF}
\begin{tabular}{llll}
\hline\hline
T / K & Q$_{\textrm{Duo}}$ & Q$_{\textrm{JPL}}$ & $\Delta$ /\% \\
\hline
300   & 7027.4727          & 6633.1886        & 5.61     \\
225   & 4694.9814          & 4491.5223        & 4.33     \\
150   & 2734.0243          & 2645.0047        & 3.26     \\
75    & 1134.4648          & 1109.3245        & 2.22     \\
37.5  & 518.8865           & 516.1226         & 0.53     \\
18.75 & 262.9106           & 262.9886         & -0.030   \\
9.375 & 139.2536           & 139.3226         & -0.050   \\
\hline
\end{tabular}
\end{table}

\begin{table*}
\caption{Uncertainty estimates (in cm$^{-1}$) are applied for the rovibronic energy levels of NaO as a function of state quantum numbers.}
\label{Table:Uncerts}
\begin{tabular}{ccccc}
\hline\hline
State                   & \multicolumn{3}{c}{X~$^2\Pi$}                                         & A~$^2\Sigma^+$\\
\hline
Case                    & $J_X < 16.5$, $v_X = 0$   & $J_X > 16.5$, $v_X = 0$           & $v_X > 0$    & all $v_A$ and $J$ levels \\
Uncertainty & 0.003                 & $0.001J_X^2 + 0.01J_X + 0.01$     & 20-40         & 10-30   \\
\hline
\end{tabular}
\end{table*}

\subsection{Final line list and absorption spectra simulation}
The \Duo~ package was used to produce a NaO line list which we call NaOUCMe. This line list is given in ExoMol format
\citep{jt584,jt810} as "*.states" file containing term values for the lowest 36~120 rovibronic levels of the $A\sim X$ complex up to $J=200.5$ and $v_{\rm X} = 29$ and a "*.trans" file consisting of the corresponding Einstein coefficients and frequencies for 4~726~137 dipole-allowed $X-X$, $A-A$ and $X-A$ transitions. In line with the updated
ExoMol standard \citep{jt810}, the "*.states" files also provide approximate uncertainties for the energies which allow the accuracy with which we predict each transition to be estimated.
For fitted regions, we recommend transition uncertainties equal to those in Table \ref{Table:RMS}. For the ground electronic state, for the case where $J < 16.5$ and $v_X = 0$, an uncertainty of $0.003~\rm{cm}^{-1}$ is recommended. Typically, energy level uncertainties scale with $J$ quadratically and the following functional form for $J > 15.5$ is recommended: $\delta E \approx 0.001J^2 - 0.01J+ 0.01$. For all other (extrapolated) cases in the X state, the {\it ab initio}  vibrational uncertainty takes precedence at 20-40~cm$^{-1}$. 

The vibrational spacing between the $v_{\rm A} = 0$ and $v_{\rm A} = 1$ states were fit to a satisfactory level of agreement (\textit{artificial line} from sec.~\ref{sec:Fitting}). However, it is recommended that this line list only be used for high resolution observational studies at long wavelengths spectral range covering pure rotational transitions. Overall energy level uncertainties are summarised in Table \ref{Table:Uncerts}.

\begin{figure}
    \includegraphics[width=\columnwidth]{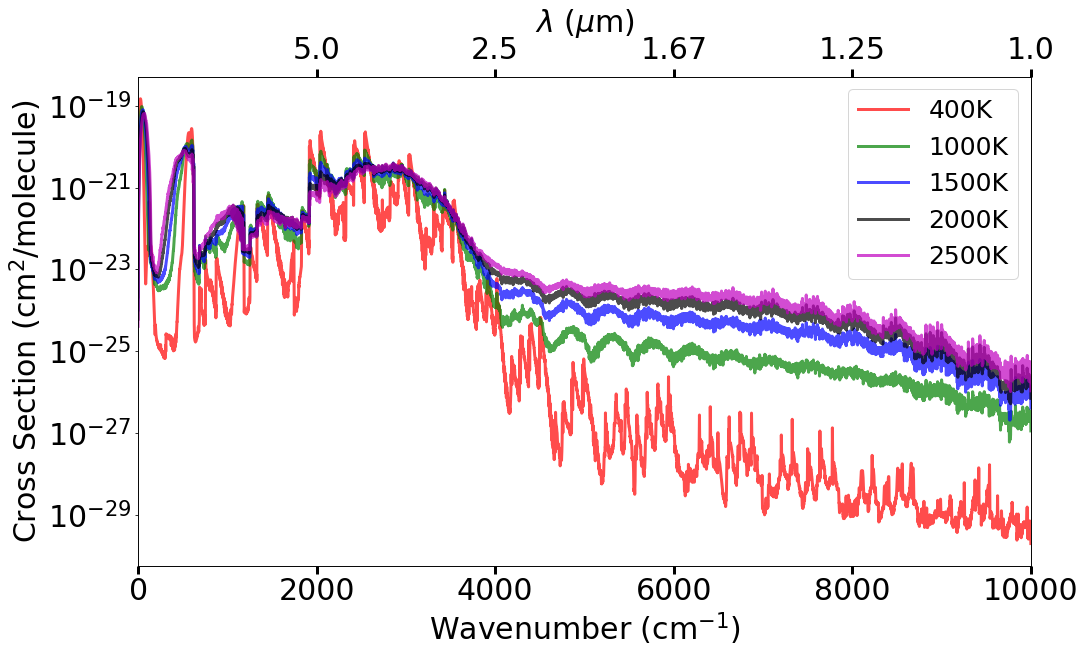}
    \caption{Spectrum of NaO at varying temperatures. Absorption cross sections were modelled with a Gaussian line profile with HWHM at 1 cm$^{-1}$ with an output on a grid with 1 \cm spacing.}
    \label{fig:CompSpec}
\end{figure}

Samples from the "*.states" and "*.trans" files are given in Tables \ref{Table:States} and \ref{Table:Trans}. These were processed in ExoCross \citep{jt708} to produce Fig. \ref{fig:Stick}, an absolute absorption spectrum for the region studied by \citet{99JoWoKo.NaO}, and Fig. \ref{fig:CompSpec}, the absorbance cross section for NaO as a function of temperature computed at a resolution of 1~cm$^{-1}$ with a half-width at half-maximum (HWHM) of 1~cm$^{-1}$.

\begin{table*}
\centering
\caption{\centering{Extract from "*.states" file produced by \textsc{Duo} for NaO. Lifetimes calculated with ExoCross, for uncertainties see Sec.\ref{sec:Conc} and Table~\ref{Table:Uncerts}}}
\label{Table:States}
\begin{tabular}{rrrrrrrrrrrrrr}
\hline\hline
\multicolumn{1}{l}{NN} &
  \multicolumn{1}{l}{Energy} &
  \multicolumn{1}{l}{gtot} &
  \multicolumn{1}{l}{$J$} &
  \multicolumn{1}{l}{uncertainty} &
  \multicolumn{1}{l}{lifetime} &
  \multicolumn{1}{l}{Land\'e} &
  \multicolumn{1}{l}{$\tau$} &
  \multicolumn{1}{l}{e/f} &
  \multicolumn{1}{l}{State} &
  \multicolumn{1}{l}{$v$} &
  \multicolumn{1}{l}{$\Lambda$} &
  \multicolumn{1}{l}{$\Sigma$} &
  \multicolumn{1}{l}{$\Omega$} \\
\hline
1  & 118.305521  & 8 & 0.5 & 0.003000   & 5.1528E+03 & 0.001493 & + & e & X2Pi    & 0 & 1 & -0.5 & 0.5 \\
2  & 704.551592  & 8 & 0.5 & 30.000000  & 2.8981E+00 & 0.001932 & + & e & X2Pi    & 1 & 1 & -0.5 & 0.5 \\
3  & 1282.294758 & 8 & 0.5 & 30.000000  & 1.4807E+00 & 0.002994 & + & e & X2Pi    & 2 & 1 & -0.5 & 0.5 \\
4  & 1850.69127  & 8 & 0.5 & 30.000000  & 1.0028E+00 & 0.012246 & + & e & X2Pi    & 3 & 1 & -0.5 & 0.5 \\
5  & 2045.653545 & 8 & 0.5 & 20.000000  & 1.0894E-01 & 1.988261 & + & e & A2Sigma & 0 & 0 & 0.5  & 0.5 \\
6  & 2409.940124 & 8 & 0.5 & 30.000000  & 6.5698E-01 & 0.041505 & + & e & X2Pi    & 4 & 1 & -0.5 & 0.5 \\
7  & 2544.609918 & 8 & 0.5 & 20.000000  & 6.4721E-02 & 1.957978 & + & e & A2Sigma & 1 & 0 & 0.5  & 0.5 \\
8  & 2960.425378 & 8 & 0.5 & 30.000000  & 3.7063E-01 & 0.112844 & + & e & X2Pi    & 5 & 1 & -0.5 & 0.5 \\
9  & 3042.140902 & 8 & 0.5 & 20.000000  & 4.9272E-02 & 1.887345 & + & e & A2Sigma & 2 & 0 & 0.5  & 0.5 \\
10 & 3501.549396 & 8 & 0.5 & 30.000000  & 1.4379E-01 & 0.388061 & + & e & X2Pi    & 6 & 1 & -0.5 & 0.5 \\
\hline
\end{tabular}
\\
\flushleft{
NN: State counting number,\\
Energy: State energy in  cm$^{-1}$,\\
gtot: Total degeneracy of the state,\\
$J$: Angular momentum quantum number,\\
uncertainty: Energy level uncertainty in cm$^{-1}$, \\
lifetime: Lifetime of the state in seconds,\\
Land\'e: Land\'e $g$ factor \citep{jt656}, \\
$\tau$: Parity, \\
e/f: Rotationless parity,\\
State: Electronic state, X2Pi or A2Sigma+\\
$v$: Vibrational quantum number,\\
$\Lambda$: Projection of electronic angular momentum,\\
$\Sigma$: Projection of electronic spin,\\
$\Omega$: $\Lambda + \Sigma$ (Projection of total electron angular momentum).
}
\end{table*}

\begin{table}
\centering
\caption{Extract from the "*.trans" file produced by \Duo\ for NaO.}
\label{Table:Trans}
\begin{tabular}{rrrr}
\hline\hline
n${_\textrm{f}}$ & n$_{\textrm{i}}$ & A$_{\textrm{fi}}$ & $\nu_{\textrm{fi}}$ \\
\hline
2097 & 2007 & 7.7832E-17 & 0.001009 \\
843  & 933  & 2.4392E-16 & 0.001035 \\
1799 & 1889 & 9.3118E-17 & 0.001043 \\
1883 & 1793 & 9.4292E-17 & 0.001055 \\
1219 & 1309 & 1.6125E-16 & 0.001066 \\
3848 & 3758 & 2.7618E-17 & 0.001137 \\
1520 & 1430 & 1.7928E-16 & 0.001148 \\
1595 & 1685 & 1.3851E-16 & 0.001153 \\
3146 & 3056 & 4.9277E-17 & 0.001177 \\
1688 & 1598 & 1.5019E-16 & 0.001179 \\
766  & 676  & 5.6326E-16 & 0.001185 \\
\hline
\end{tabular}
\\
\flushleft{
n$_{\textrm{f}}$: Lower state counting number\\
n$_{\textrm{i}}$: Upper state counting number\\
A$_{\textrm{fi}}$: Einstein A-coefficient in s$^{-1}$ \\
$\nu_{\textrm{fi}}$: Transition wavenumber in cm$^{-1}$\\
}
\end{table}

\subsection{Comparison with the experimental X$\to$ A spectra} \label{subsec:CompExp}

\begin{figure}
    \includegraphics[width=\columnwidth]{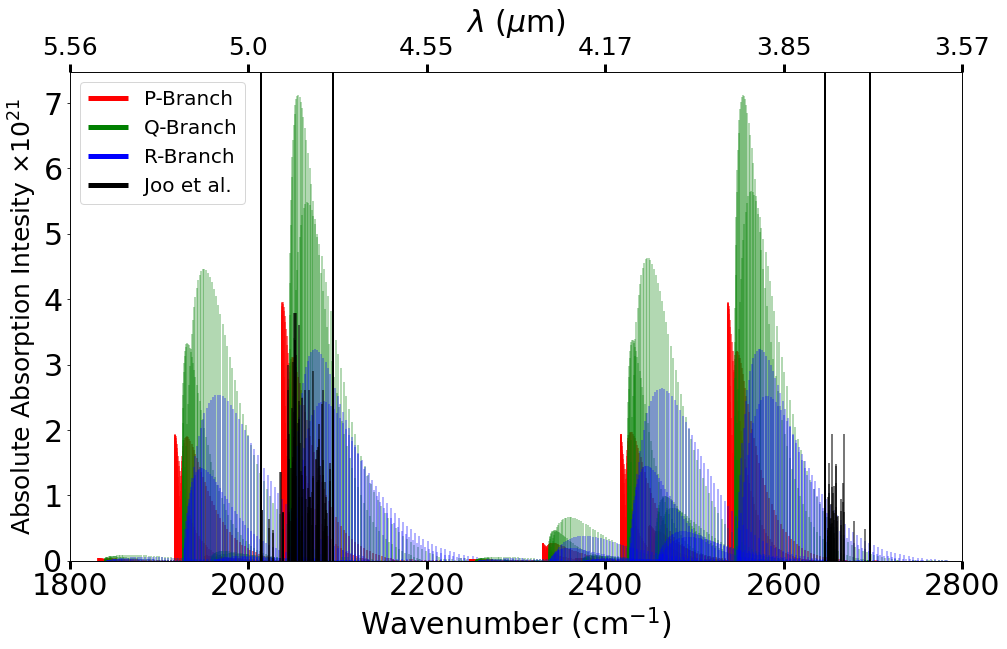}
    \caption{Absolute absorption intensity (arbitrary units) spectrum of NaO computed at $T=400$~K. Black lines represent the experimental transitions containing (from left to right) the observed A~$^2\Sigma^+(0,1) \leftarrow$ X~$^2\Pi(0)$ transition bands from \citet{99JoWoKo.NaO}. Black vertical lines indicate start and end of the observed band transition regions.}
    \label{fig:Stick}
\end{figure}

The experimental A~$^2\Sigma^+(0,1) \leftarrow$ X~$^2\Pi(0)$ absorption bands were observed by \citet{99JoWoKo.NaO} but their measurements were only sensitive in the very narrow spectral regions: 2015--2095~cm$^{-1}$ and 2646--2697~cm$^{-1}$. The lines in these regions were assigned to the $\Delta v = 0$ and $\Delta v = +1$ bands respectively. An absolute intensity absorption spectrum was calculated as a stick spectrum to compare to this, see Fig. \ref{fig:Stick}. The calculated absorption lines result in a line profile significantly broader than those visible from measurements of \citet{99JoWoKo.NaO}. The observed spectrum (relative intensities)  has been scaled by matching the intensity of the strongest line [A $(J'=7.5, v' = 0) \leftarrow$ {X} $(J''=7.5, v''=0,\Omega''=-1.5)$] from experiment to that of its conjugate calculated line; it can be seen that the observed $\Delta v = +1$ transitions lie on the tail end, $R$-branch of the calculated absorption spectrum with a matching downward slope. 

We note that  \citet{99JoWoKo.NaO} used their measurements to determine the vibronic term value for the A(0)--X(0) band, $T_0$(A), of 1992.905~cm$^{-1}$. Using approximate $J=0$ zero-point energies (ZPEs) generated by \Duo, and $T_0({\rm A}) = T_{\rm e}({\rm A}) - {\rm ZPE}({\rm X}) + {\rm ZPE}({\rm A})$ shows that our $T_0(A)$ from our deperturbation model is 1982.3~cm$^{-1}$, which is less than that of \citet{99JoWoKo.NaO} by 10~cm$^{-1}$. An attempt was made to reproduce the experimental $T_0$(A) value above by manually adjusting $T_{\rm e}$(A) and refitting. However, this results in significantly worse fits in the transitions (around an order of magnitude) without improving agreement of $T_0$(A) with experiment.

\section{Conclusions}\label{sec:Conc}
An optical line list called NaOUCMe is presented for the alkali monoxide radical, NaO covering transitions in the 0--10000~cm$^{-1}$ energy range up to $J=200.5$ and $v_X = 29$ with applicability up to the equilibrium temperature $T=2500$~K. The experimental lines for pure rotational transitions in the ground electronic state \citep{89YaFuHi.NaO} and rovibronic A~$^2\Sigma^+(0,1) \leftarrow$ X~$^2\Pi(0)$ transitions  \citep{99JoWoKo.NaO} were used to refine the deperturbation model applied.  An average RMS of $\sim 0.006$ \cm\ was achieved for the \X\ state rotational transitions and $0.35~\rm{cm}^{-1}$ for the \A\ $\leftarrow$ \X\ transitions, with $\sim 0.023~\rm{cm}^{-1}$ RMS for the A$(v' = 0) \leftarrow$ X$(v'' = 0)$ transitions and $\sim 0.680~\rm{cm}^{-1}$ RMS for the A$(v' = 1, J' = 0.5) \leftarrow$ A$(v'' = 0, J'' = 0.5)$ \textit{artificial line} (see sec. \ref{sec:Fitting}). 
Uncertainties of the reproduced and predicted lines rapidly increase as the vibrational and rotational quantum numbers increase for both electronic states, and, hence, further experimental data on higher excited transitions is required to create a more robust line list. 

The success of the CaO line list \citep{16YuBlAs.CaO} in a spectrum assignment at 29~km above the Earth's surface by \citet{18BeBoSa.CaO} is indicative of potential importance of s-block metal monoxides like NaO. NaO similarly is a molecule that can be studied in a terrestrial environment at approximately 90~km above the Earth's surface \citep{05SlCoHuSa} and has been recently studied as a transient species in airglow by \citet{09SaSeChNa,10SaSeChSr} and \citet{12PlOeMiSa}. As such, we strongly encourage further experimental study into NaO for refinement of the spectroscopic model and production of a higher accurate line list.

\section*{Acknowledgements}
We thank Dr. A. A. Berezhnoy for bringing the problem to our attention and Prof. I. A. Osterman for taking part in the preliminary {\it ab initio} calculations. 
This work was supported by ERC Advanced Investigator Project 883830 and by UK STFC under grant ST/R000476/1. A.V.S. is grateful for the support by the Russian Science Foundation (RSF), Grant No.18-13-00269.

\section*{Data Availability}

The \textsc{Duo} model input file and calculated partition function file given
 as a supplementary materials to this article.
The NaO NaOUCMe states and transition files of  can be downloaded from
\href{www.exomol.com}{www.exomol.com} and the CDS data centre
\href{http://cdsarc.u-strasbg.fr}{cdsarc.u-strasbg.fr}. The open access  programs \textsc{ExoCross} and \textsc{Duo} are  available from \href{https://github.com/exomol}{github.com/exomol}.



\bibliographystyle{mnras}
\bibliography{AbInitio,References,journals_astro,jtj,NaO,programs,CaO,FeO,partition,methods, Books} 




\bsp	
\label{lastpage}
\end{document}